\documentstyle[12pt]{article}
\begin{document}
\begin{center}
{\Large On the necessity of non-Riemannian acoustic spacetime in fluids with vorticity}
\vspace{1cm}

\noindent
L.C. Garcia de Andrade\footnote{Departamento de Fisica Teorica-Instituto de F\'{\i}sica , Universidade do Estado do Rio de Janeiro-UERJ, Rua S\~{a}o
Francisco Xavier 524, Rio de Janeiro Brasil.E-mail:garcia@dft.if.uerj.br}
\end{center}
\vspace{2cm}
\begin{center}
{\Large Abstract}
\end{center}
\vspace{0.5cm}
The necessity of a newly proposed (PRD 70 (2004) 64004) non-Riemannian acoustic spacetime structure called acoustic torsion of sound wave equation in fluids with vorticity are discussed. It is shown that this structure , although not always necessary is present in fluids with vorticity even when the perturbation is rotational. This can be done by solving the Bergliaffa et al (Physica D (2004)) gauge invariant equations for sound, superposed to a general background flow, needs to support a non-Riemannian acoustic geometry in effective spacetime. Bergliaffa et al have previously shown that a Riemannian structure cannot be associated to this gauge invariant general system. 
\newpage
\section{Introduction}
In recent papers \cite{1,2}, Riemannian and non-Riemannian acoustic metric and acoustic torsion structures have been considered. In the first reference Bergliaffa et al have shown that a Riemannian acoustic metric may be written for rotational flows as long as the perturbation to this rotational flow is potential. More recently we have shown that the non-Riemannian structure called acoustic torsion may also exist in this same case. Therefore the in this Brief report, based on Bergliaffa et \cite{2} formalism where the Riemannian acoustic structure is not enough to describe the effective acoustic geometry, the necessity of acoustic torsion on a particular solution of their gauge invariant system of sound equations for  rotational perturbations implies a natural non-Riemannian geometrical interpretation. Unfortunately as point out by the authors of reference \cite{2}, no general Riemannian or non-Riemannian structure, so far has been able to provide a general solution to the problem of acoustic spacetimes of general vorticity in general background flows. This brief report is organised as follows : In the section 2 a particular solution of gauge invariant equations for sound in fluids with vorticity is obtained where approximations are used to fit the non-Riemannian structure in the vortex acoustics spacetime. 
\section{The need for non-Riemannian acoustic structure in flows with vorticity}
Several papers on acoustic geometry have been released in the last years , all of them considering Riemannian acoustic geometry as can be seen in the review given in refrence \cite{3}. A couple of exceptions are given by Barcelo et al \cite{4} where they deal with the non-Riemannian Finslerian geometry. Although not considered explicitely by the authors in general Finsler spaces possesses three kinds of torsions. In this section we addopt a simpler approach considering as in reference \cite{1}, a Cartan acoustic torsion analogous to the one given in Einstein-Cartan gravity alternative to Einstein's general relativity. In this case the wave equation considered in the effective spacetime would be given by the generalised wave equation in Riemann-Cartan spacetime given by
\begin{equation}
{\Box}{\psi}=-{K^{i}}{\partial}_{i}{\psi}
\label{1}
\end{equation}
Here ${\Box}$ represents the Riemannian D'Alembertian and $K^{i}$ $(i=0,1,2,3)$ is the torsion trace. Starting from the Clebsch representation of the  velocity field $\vec{v}={\nabla}{\phi}+{\beta}{\nabla}{\gamma}$, where ${\beta}$ and ${\gamma}$ are the Clebsch potentials dependent of time and space  Bergliaffa et al \cite{2} were able to obtain a gauge invariant system of equations
\begin{equation}
\frac{d}{dt}(\frac{1}{c^{2}}\frac{d{\psi}_{1}}{dt})=\frac{1}{{\rho}_{0}}{\nabla}({\rho}_{0}({\nabla}{\psi}_{1}+{\epsilon}_{1}))
\label{2}
\end{equation}
and
\begin{equation}
\frac{d}{dt}{\epsilon}_{1}= {\nabla}{\psi}_{1}{\times}\vec{\omega}_{0}-({\epsilon}_{1}.{\nabla})\vec{v_{0}}
\label{3}
\end{equation}
where $c^{2}$ is the square of local speed of sound, and quantities with zero indices represent background flow quantities while quantities with one indices represent the perturbed quantities over the background flow. The perturbed velocity can be expressed as 
\begin{equation}
\vec{v_{1}}= {\nabla}{\psi}_{1}+{\epsilon}_{1}
\label{4}
\end{equation}
which shows clearly the ${\epsilon}_{1}$ breaks the potential flow perturbation of the first term in the perturbed speed $\vec{v_{1}}$. The quantity 
\begin{equation}
{\epsilon}_{1}={x}_{1}{\times}{\omega}_{0} 
\label{5}
\end{equation}
gives the product of the displacement field  of a material particle due to the sound wave by the background rotation frequency. This quantity can also be expressed in terms of the Clebsch potentials by \cite{1}
\begin{equation}
{\epsilon}_{1}={\beta}_{1}{\nabla}{\gamma}_{0}-{\gamma}_{1}{\nabla}{\beta}_{0}
\label{6}
\end{equation}
If one  assumes ,as in reference \cite{1}, that we can ignore the quantity ${\epsilon}_{1}$ we are left with the  gauge invariant system reduces to the Riemannian acoustic the wave equation 
\begin{equation}
(\frac{{\partial}}{{\partial}t}+\vec{v}_{0}.{\nabla})\frac{1}{c^{2}}(\frac{{\partial}}{{\partial}t}+\vec{v}_{0}.{\nabla}){\psi}_{1}=\frac{1}{{\rho}_{0}}{\nabla}.({\rho}_{0}{\nabla}{\psi}_{1})
\label{7}
\end{equation}
This equation yields the Unruh's acoustic metric \cite{4} 
\begin{equation}
ds^{2}=\frac{{\rho}_{0}}{c}[c^{2}dt^{2}-{\delta}_{ab}(dx^{a}-{v^{a}}_{0}dt)(dx^{b}-{v^{b}}_{0}dt)]
\label{8}
\end{equation}
where $(a,b=1,2,3)$. Now considering the general case of the rotational fluid it is possible to  Riemannian wave equation yields
\begin{equation}
(\frac{{\partial}}{{\partial}t}+\vec{v}_{0}.{\nabla})\frac{1}{c^{2}}(\frac{{\partial}}{{\partial}t}+\vec{v}_{0}.{\nabla}){\psi}_{1}=\frac{1}{{\rho}_{0}}{\nabla}.({\rho}_{0}[{\nabla}{\psi}_{1}+{\epsilon}_{1}])
\label{9}
\end{equation}
Expanding this equation ,now it is possible to show that the wave equation breaks into a Riemannian plus a non-Riemannian part where an acoustic torsion vector can be defined. This equation can be written as 
\begin{equation}
(\frac{{\partial}}{{\partial}t}+\vec{v}_{0}.{\nabla})\frac{1}{c^{2}}(\frac{{\partial}}{{\partial}t}+\vec{v}_{0}.{\nabla}){\psi}_{1}=\frac{1}{{\rho}_{0}}{\nabla}.({\rho}_{0}{\nabla}{\psi}_{1})+\frac{1}{{\rho}_{0}}{\nabla}.({\rho}_{0}\vec{{\epsilon}_{1}})
\label{10}
\end{equation}
To be able to find out the particular solution to the gauge invariant system we impose the restriction on the flow given by
\begin{equation}
(\vec{{\epsilon}_{1}}.{\nabla}){\vec{v}_{0}}=0
\label{11}
\end{equation}
This simplifies equation (\ref{3}) to
\begin{equation}
\frac{d}{dt}{\epsilon}_{1}= {\nabla}{\psi}_{1}{\times}\vec{\omega}_{0}
\label{12}
\end{equation}
Under certain assumptions this equation, now can be easily integrated to yield
\begin{equation}
{\epsilon}_{1}= \int{{\nabla}{\psi}_{1}{\times}\vec{\omega}_{0}dt}
\label{13}
\end{equation}
By considering that ${\nabla}{\psi}_{1}$ vaies very slowly in time ,which is a reasonable assumption in hydrodynamics, one obtains the following expression for ${\epsilon}_{1}$
\begin{equation}
{\epsilon}_{1}= {\nabla}{\psi}_{1}{\times}\vec{{\theta}_{0}}
\label{14}
\end{equation}
where the angular displacement $\vec{{\theta}_{0}}$ is given by
\begin{equation}
\vec{{\theta}_{0}}= \int{\vec{\omega}_{0}dt}
\label{15}
\end{equation}
substitution of expression {\ref{14}) into the other equation of the gauge invariant system for the sound in the fluid with vorticity ,one obtains the non-Riemannian $(NR)$ term of the sound wave equation as
\begin{equation}
NR=[{\nabla}-(\frac{{\nabla}{{\rho}_{0}}}{{\rho}_{0}}]{\times}\vec{{\theta}_{0}}.{\nabla}{\psi}_{1}
\label{16}
\end{equation}
By comparison with the non-Riemannian wave equation (\ref{1}), and consider the follwoing structure to the torsion vector $(K^{0}=0)$ and only the spatial part $\vec{K}$ surviving we obtain
\begin{equation}
\vec{K}= [{\nabla}-\frac{{\nabla}{\rho}_{0}}{{\rho}_{0}}]{\times}\int{\vec{{\omega}_{0}}dt}
\label{17}
\end{equation}
This expression shows that the acoustic torsion is expressed in terms of the background rotation frequency. This result is similar to the one obtained in reference \cite{2}, however, the main and important difference here is that the perturbation flow now is rotational, contrary to the previous case. This physical situation solves also a problem in the acoustic perturbation physics, since now there is no contamination of the irrotational perturbation by the rotational background flow since both are vorticity flows.
\section{Conclusions}
We solved a recent controversy \cite{6} on the existence and necessity of the non-Riemannian structure of acoustic spacetime in fluids with vorticity by solving the gauge invariant system of wave equations in general background flow with vorticity. This solution is shown to support a non-Riemannian structure called acoustic torsion.

\end{document}